\documentclass[review]{elsarticle}

\usepackage{hyperref}
\usepackage{tikz}
\usepackage{amsmath}
\usetikzlibrary{arrows}
\usepackage{subcaption}

\journal{Journal of \LaTeX\ Templates}

\begin{document}

\begin{frontmatter}

\title{Atmospheric heat redistribution effect on Emission spectra of Hot-Jupiters}

\author[mymainaddress,mysecondaryaddress]{Soumya Sengupta}

\cortext[mycorrespondingauthor]{Corresponding author}
\ead{soumya.s@iiap.res.in}

\author[mymainaddress]{Sujan Sengupta}



\address[mymainaddress]{Indian Institute of Astrophysics, Koramangala 2nd Block, Sarjapura Road, Bangalore 560034, India}
\address[mysecondaryaddress]{Pondicherry University, R.V. Nagar, Kalapet, 605014, Puducherry, India}


\begin{abstract}
Hot Jupiters are the most studied and easily detectable exoplanets for transit observations. However, the correlation between the atmospheric flow and the emission spectra of such planets is still not understood.
Due to huge day-night temperature contrast in {hot Jupiter},  the thermal redistribution through atmospheric circulation has a significant impact on the vertical temperature-pressure structure and on the emission spectra. In the present work, we aim to study the variation of the  temperature-pressure profiles and the emission spectra of such planets due to different amounts of atmospheric heat redistribution.
For this purpose, we first derive an analytical relation between the heat redistribution parameter $\rm f$ and the emitted flux from the uppermost atmospheric layers of {hot Jupiter}. We adopt the three possible values of  $\rm f$ under isotropic approximation as $\frac{1}{4}, \frac{1}{2}$ and $\frac{2}{3}$ for full-redistribution, semi-redistribution and no-redistribution cases respectively and calculate the corresponding temperature-pressure profiles and the emission spectra. Next, we model the emission spectra for different values of $\rm f$ by numerically solving the radiative transfer equations using the discrete space theory formalism. We demonstrate that the atmospheric temperature-pressure profiles and the emission spectra  both are susceptible to the values of the heat redistribution function. A reduction in the heat redistribution yields a thermal inversion in the temperature-pressure profiles and hence {increases} the amount of emission flux. Finally, we revisits the {hot Jupiter} XO-1b temperature-pressure profile degeneracy case and show that a non-inversion temperature-pressure profile best explains this planet's observed dayside emission spectra.  
\end{abstract}


\end{frontmatter}


\section{Introduction}\label{intro}

Since the detection of the first Jupiter-like extrasolar planet 51-Pegasi-b around a solar type star \citep{mayor1995jupiter}, the giant planets remain the most observed and investigated exoplanets till date. For transit observations, {hot Jupiter}s are the first as well as the most easily detectable planets \citep{mazeh2000spectroscopic}, \citep{knutson2007map} among a large variety of exoplanets discovered. The atmospheric temperature-pressure profiles of {hot Jupiter}s have been modeled using the radiative equilibrium conditions both analytically \citep{hansen2008absorption,guillot2010radiative}, as well as numerically \citep{parmentier2014non, parmentier2015non}. Using these modelled temperature-pressure profiles and the atmospheric chemical compositions, the atmospheric spectra of such exoplanets can be obtained by solving the radiative transfer equations
 {(e.g. \citep{burrows2008optical,tinetti2013spectroscopy,molliere2019petitradtrans}) with different techniques such as two-stream approximation \citep{malik2017helios,drummond2018effect}, Feautrier method \cite{gandhi2017genesis}, Discrete Space Theory\cite{sengupta2009multiple} etc.}
 Ultimately, by comparing those synthetic spectra with the extracted spectra from transit observations, one can retrieve the atmospheric chemical compositions {\citep{tinetti2007water, swain2008presence,madhusudhan2014h2o,fraine2014water,kreidberg2014clouds,barstow2016consistent,benneke2012atmospheric}} and the temperature-pressure profiles {\citep{nikolov2018hubble, madhusudhan2009temperature, tinetti2010probing, brogi2019retrieving,zhang2019forward}}.{The atmospheric retrieval of a huge number of Gaseous and hot Jupiter planetary atmospheres are reported in \cite{fisher2018retrieval,tsiaras2018population}.}

 Three types of spectra that can be obtained during transit and eclipse observations, e.g., transmission, reflection and emission spectra \citep{tinetti2013spectroscopy} are studied extensively (e.g.{ \cite{sengupta2020optical,chakrabarty2020effects,waldmann2015tau,molliere2019petitradtrans,kempton2017exo,batalha2019exoplanet,gibson2020detection,waldmann2013signals}}). Among all these, only the planetary emission spectra observed during  secondary eclipse observations carry the full imprint of the atmospheric temperature-pressure profiles {\citep{tinetti2013spectroscopy,gandhi2019hydra,shulyak2019remote,madhusudhan2011high,greene2016characterizing,blecic2017implications,line2013systematic}}, whereas reflection and transmission spectra are sensitive to the upper atmosphere only \citep{sengupta2020optical}.
 {The emission}
 spectra  carry the compositional signature of the atmospheric layers in terms of absorption and scattering as well as the temperature of each atmospheric layer \citep{sengupta2021effects,sengupta2022atmospheric}. Due to the tidal locking of the {hot Jupiter}s with its host star, there is a huge difference in the atmospheric temperature at the permanent day-side and the permanent night-side of the planet. This temperature difference 
 {gives rise to a pressure gradient which results to the horizontal atmospheric flow \cite{knutson2007map}. Also the hot Jupiter's small rotation period effects the horizontal flow geometry in terms of coriolis force which is very much different than the solar system giants \cite{showman2007atmospheric}.}
Now the stellar irradiation flux has a direct effect on planetary spectra through re-emission \citep{chakrabarty2020effects} and in heat redistribution through equatorial-jet circulation \citep{hammond2020equatorial}. It was pedagogically shown in \cite{hansen2008absorption} that the planetary atmospheric heat redistribution has a direct effect on the temperature-pressure profile as well as in the emission spectra. Recently, \cite{komacek2019temporal} demonstrated that due to large scale equatorial waves generated by prominent day-night temperature contrast, there is a temporal variation in the secondary eclipse depth (~2\% locally) and the temperature-pressure profiles. A number of studies of atmospheric heat redistribution has been done using different available mechanisms (see \cite{seager2010exoplanets, heng2015atmospheric, showman2020atmospheric} and the references therein). Initially the shallow water model is used to explain the atmospheric heat redistribution in {hot Jupiter}s \citep{perez2013atmospheric}. Recently, \cite{tan2019atmospheric} investigated  the heat transfer from dayside to nightside in ultra hot jupiters by considering the General Circulation Model with an effect of molecular hydrogen dissociation in dayside and atomic hydrogen recombination in nightside. Hence each vertical atmospheric layer in the dayside atmosphere is cooled due to the heat transfer from substellar to anti-stellar side of the planet.

Although, different kind of atmospheric spectra of {hot Jupiter}s has been studied, the effect of atmospheric heat redistribution on {hot Jupiter}'s atmosphere remains unresolved.  Thus, to get the full realization of the atmospheric heat redistribution, one needs to study the temperature-pressure profiles as well as the emission spectra for different degree of atmospheric heat redistribution. In this paper, we present the effect of day to night side atmospheric heat redistribution of {hot Jupiter}s. We however, do not address the possible mechanisms of heat redistribution. The effect is analyzed in terms of temperature-pressure profile and day-side emission spectrum as suggested by \cite{hansen2008absorption}. We use the analytical formulation of temperature-pressure profiles presented by \cite{guillot2010radiative} to derive the analytical relation of redistribution parameter with the emission profiles. Using these profiles and solar abundance composition in the atmosphere, the absorption as well as the scattering co-efficients is estimated by using the publicly available software package Exo-Transmit \citep{kempton2017exo, freedman2008line, freedman2014gaseous, lupu2014atmospheres}.  Finally, the modelled temperature-pressure profile, absorption and scattering co-efficients are used to calculate the dayside emission spectra numerically by using discrete space theory formalism described in \cite{sengupta2009multiple}, \cite{sengupta2020optical}. Thus, from the emission spectra we infer the role of heat redistribution in the atmosphere of hot-juipters.

In section~\ref{model}, we present the analytical derivations of the expression that describe the explicit dependence of atmospheric thermal profile as well as the dayside emission spectra on the redistribution parameter $\rm f$.  The numerical method for solving the 1-D radiative transfer equations is described and the solutions are validated in section~\ref{methodology}. The resulting effects of redistribution parameter on the temperature-pressure profile and the dayside emission spectra are presented in section~\ref{Results}. Finally we discuss our results and conclude this work in the last section.

\section{Theoretical Models}\label{model}
A hot close-in gas giant planet is tidally locked to its host star. When the starlight is irradiated on the  atmosphere, then two phenomena can occur: Either the heat can be absorbed and redistributed in the atmosphere or the heat is reradiated immediately from the atmosphere. 

The analytical expression for the atmospheric temperature
{T} 
under radiative equilibrium is given by \cite{guillot2010radiative} 
\begin{equation}\label{TP for collimated}
T^4 = \frac{3T_{int}^4}{4}\left(\frac{2}{3}+ \tau\right) + \frac{3T_{irr}^4}{4}\mu_0\left[\frac{2}{3}+\frac{\mu_0}{\gamma}+\left(\frac{\gamma}{3\mu_0}-\frac{\mu_0}{\gamma}\right)e^{-\frac{\gamma\tau}{\mu_0}}\right]
\end{equation}
where,$T_{int}$ is the internal temperature of the planet, $T_{irr}$ is temperature due to the flux irradiated on the planetary atmosphere along the direction cosine $\mu_0$.
{Here $\mu_0$ is the direction cosine of the angle made by the irradiated radiation with the outward normal of the atmospheric layer \cite{guillot2010radiative}.}
$\gamma$ is the ratio between the optical and the infrared absorption co-efficients, i.e. $\kappa_{vis}/\kappa_{inf}$ and $\tau$ is the optical depth defined in terms of pressure (P), density ($\rho$) and constant surface gravity g as, $d\tau = \frac{dP{\kappa_{inf}}}{g}$ \citep{guillot2010radiative}.

Thus the mean intensity expression can be obtained by multiplying equation \eqref{TP for collimated} by $\frac{\sigma}{\pi}$, where $\sigma$ is the Stefan-Boltzmann constant, on either side as,
\begin{equation}\label{mean intensity for collimated}
J = \frac{3}{4}F_{int}\left(\frac{2}{3}+ \tau\right) + \frac{3}{4}F_{irr}\mu_0\left[\frac{2}{3}+\frac{\mu_0}{\gamma}+\left(\frac{\gamma}{3\mu_0}-\frac{\mu_0}{\gamma}\right)e^{-\frac{\gamma\tau}{\mu_0}}\right]
\end{equation}
where, we replaced $\sigma T_{int}^4/\pi$  and $\sigma T_{irr}^4/\pi$ by internal flux $F_{int}$ and irradiated flux $F_{irr}$ respectively and make use the fact that $J=\frac{\sigma T^4}{\pi}$ \citep{Chandrasekhar}.

Under radiative equilibrium condition, the specific intensity at $\tau = 0$ and the mean intensity relation can be
{written} 
as \citep{hansen2008absorption},
\begin{equation}\label{I and J rltn}
I(\tau=0,\mu,\mu_0) = \int_0^\infty J(t) e^{-\frac{t}{\mu}} \frac{dt}{\mu}
\end{equation}

Now solving equation \eqref{I and J rltn} by using equation \eqref{mean intensity for collimated} we get,
\begin{equation}\label{I for collimated}
I(0,\mu,\mu_0) = \frac{3F_{int}}{4}(\frac{2}{3}+ \mu) + \frac{3F_{irr}}{4}\mu_0\big[\frac{2}{3}+\frac{\mu_0}{\gamma}+\big(\frac{\gamma}{3\mu_0}-\frac{\mu_0}{\gamma})\frac{1}{1+\gamma \mu/\mu_0}]
\end{equation}
This is the flux emerging out from the  uppermost atmospheric layer of the planet where $\tau=0$. For secondary eclipse observation, the total observed flux is the radiation coming from the substellar point of the planet at full phase with $\mu=\mu_0$. Taking the integral over this phase and following the notations of \cite{hansen2008absorption},  we can write the full emerging flux as,
\begin{equation}\label{full secondary eclipse flux 1}
F_{full} = 2\int_0^1 \mu_0 I(0,\mu_0,\mu_0)d\mu_0
= F_{int}
+\frac{3F_{irr}}{2} \left[\frac{2}{9} +\frac{1}{4\gamma}+\left(\frac{\gamma}{6}-\frac{1}{4\gamma}\right)\frac{1}{1+\gamma}\right]
\end{equation}

{in principle,}
the comparison of the flux observed during secondary eclipse to the model flux $F_{full}$ can tell us whether the energy is redistributed from substellar side to anti-stellar side of the planet, because the observed emitted flux from the substellar side in that case would  be less than that expected for no-redistribution model. Thus dividing the re-radiation term i.e. the second term in the right hand side of equation \eqref{full secondary eclipse flux 1} by the irradiated flux $F_{irr}$ we get an effective redistribution factor,
\begin{equation}\label{effective f}
f_{eff} = \frac{3}{2} \left[\frac{2}{9} +\frac{1}{4\gamma}+\left(\frac{\gamma}{6}-\frac{1}{4\gamma}\right)\frac{1}{1+\gamma}\right]
\end{equation}

Similarly, the irradiated flux expressed by equation \eqref{full secondary eclipse flux 1} can also be written in terms of the zero albedo equilibrium flux at planetary atmosphere (i.e. $F_{eq0}$). To do that we consider the average temperature-pressure profile with $\mu_0 = 1/\sqrt{3}$ as shown in \cite{guillot2010radiative} and can be denoted by a particular term $\mu_*$. This is known as isotropic approximation case and has been considered by \cite{parmentier2015non} using the following relations:
\begin{equation}\label{isotropic condition}
\mu_* = \frac{1}{\sqrt{3}}
T_{\mu_*}^4 = \mu_* T_{irr}^4 =\frac{1}{\sqrt{3}} T_{irr}^4 = (1-A_B)4f T_{eq0}^4
\end{equation}
where, $A_B$ is the Bond Albedo, $T_{eq0}$ is the equilibrium temperature at zero albedo and $\rm f$ is the atmospheric redistribution parameter. We emphasize here that $\rm f$ and $f_{eff}$ are two completely different parameters. $\rm f$ is the redistribution parameter that tells how much day to night side horizontal atmospheric flow occurs, whereas $f_{eff}$ is a function  of $\gamma$ and represents the variation of flux absorbed at different atmospheric depth. 
Thus, for isotropic approximation $F_{irr}$ of equation \eqref{full secondary eclipse flux 1} can be written as,
\begin{equation}\label{F_irr in isotropic}
F_{irr} = \frac{\sigma}{\pi} T_{irr}^4
= 4\sqrt{3}(1-A_B)f F_{eq0}
\end{equation}
where  $F_{eq0} = \frac{\sigma}{\pi}T_{eq0}^4$

Thus considering equation \eqref{effective f}, \eqref{isotropic condition} and \eqref{F_irr in isotropic} we can write equation  \eqref{full secondary eclipse flux 1} as,
\begin{equation}\label{full secondary eclipse flux 2}
F_{full} = F_{int} + 4\sqrt{3}(1-A_B)f_{eff} f F_{eq0}
\end{equation}

Equation \eqref{full secondary eclipse flux 2} represents total emergent flux in case of secondary eclipse emission spectra. In this work we study the particular case for fixed internal and equilibrium temperature as well as for fixed albedo. Thus, the total flux $F_{full}$ may vary depending on the parameters $f_{eff}$ and $\rm f$ only.

\begin{figure}[ht!]
\begin{center}
\includegraphics[width = 10cm,height=7cm]{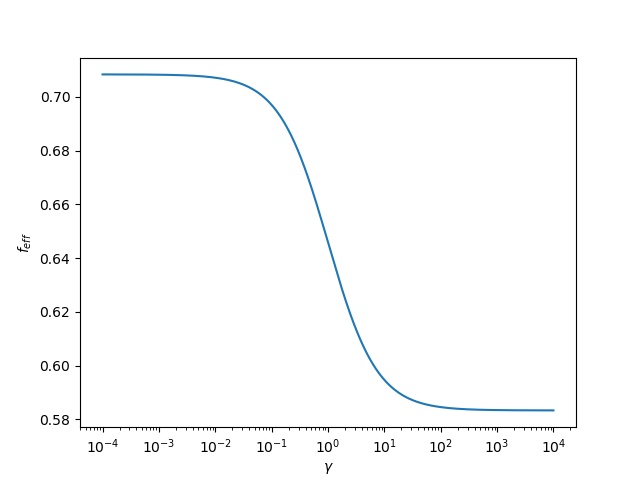}
\end{center}
\caption{ Variation of the function $f_{eff}$  with $\gamma$, the ratio between the optical and the infrared absorption co-efficients. }
\label{redistribution_vs_gamma}
\end{figure}

The explicit form of $f_{eff}$, is given in Equation \eqref{effective f} and its variation with $\gamma$ is shown in figure \ref{redistribution_vs_gamma}. These are equivalent to that obtained by \cite{hansen2008absorption} except some different boundary conditions used. The figure shows that $f_{eff}$ saturates both at large and small $\gamma$. Also it reveals that $f_{eff}$ is very insensitive to $\gamma$. The values of $f_{eff}$ saturate at $f_{eff}\to 0.7083$ for $\gamma\to 0$ and $f_{eff}\to 0.583$ for $\gamma\to\infty$ as derived using python package "simpy" \citep{meurer2017sympy}  available in public domain\footnote{\href{https://www.sympy.org/en/index.html}{https://www.sympy.org/en/index.html}} .  This insensitivity of $f_{eff}$ to $\gamma$ implies that the variation in the emission spectra during the secondary eclipse is effectively due to the atmospheric heat redistribution factor $\rm f$.  It also suggests that the emission spectra almost remain unaffected by the atmospheric depth at which the irradiated energy penetrates. Therefore the degree of thermal redistribution in the planetary atmosphere can be 
{estimated}
by comparing the observed flux during the secondary eclipse with model spectra calculated by using different heat redistribution parameters
{f in the range 0.1 to 0.9.} 
\subsection{Values of the redistribution parameter for isotropic approximation}\label{f for isotropic approximation}

\begin{figure}[ht!]
\begin{center}
\begin{tikzpicture}
\draw (0,0) circle(1.5cm);
\fill[black!70] (0,0) circle(1.5cm);
\fill[white] (0,0) -- (-90:1.5cm) arc (-90:90:1.5cm) -- cycle;
\draw (0,0) -- (1.06,1.06);
\draw node at (0.5,0.1) {$R_p$};
\draw [dashed](4,0) ellipse(0.5cm and 1.5cm);
\draw [dashed](4,0) -- (4,1.5);
\draw node at (4.3,0.5) {$R_p$};
\draw[red] (9,3)node[right]{star} arc(150:200:7cm);
\draw[thick,<-] (1.6,0) -- (8,0);
\draw[thick,<-] (0.5,1.5) -- (8.4,1.5);
\draw[thick,<-] (0.5,-1.5) -- (8,-1.5);
\draw[thick,<->] (0,-2.1) -- (8,-2.1);
\draw node at (4,-2.3) {a};
\draw [thick,->] (0.1,1.5)--(0.4,2);
\draw [thick,->] (1.25,1.05)--(1.7,1.4);
\draw [thick,->] (1.4,0.5)--(2.0,0.8);
\draw [thick,->] (1.4,-0.5)--(2.0,-0.8);
\draw [thick,->] (1.25,-1.05)--(1.7,-1.4);
\draw [thick,->] (0.1,-1.5)--(0.4,-2);
\draw[thick,green, ->] (0.7,-0.3) arc (0:-190:0.7);
\draw[thick,green, <-] (0.7,0.3) arc (0:190:0.7);
\end{tikzpicture}
\caption{The isotropic irradiation approximation. The starlight incidents on the dayside of the planet isotropically through the area $\pi R_p^2$. Due to atmospheric heat redistribution from dayside to nightside (shown in green arrow), the redistribution parameter is altered.  The energy emitted from the dayside of a hot Jupiter decreases with the increase in the heat redistribution.}
\label{isotropic irradiation approximation}
\end{center}
\end{figure}
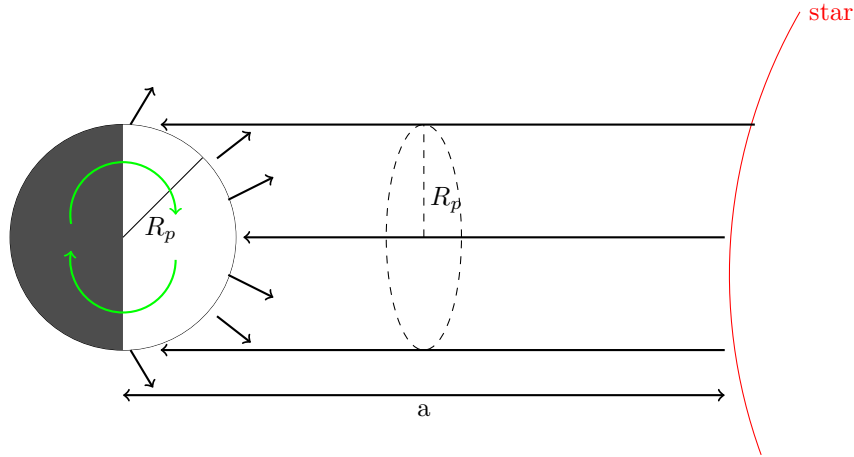

{As} the close-in {hot Jupiter}s are tidally locked with their parent star, they have 
{permanent}
day and night sides. Under such circumstance, the irradiation can be treated by isotropic approximation \citep{guillot2010radiative}. In this case the irradiation flux is assumed to be incident on
{the planetary disk of}
circular  area $\pi R_p^2$ as shown in fig. \ref{isotropic irradiation approximation},
{where $R_p$ represents the planetary radius.}
Before the energy is re-emitted from the planet, the irradiated energy is redistributed on the planetary atmosphere in three possible ways. The redistribution parameter $\rm f $ can be defined as the ratio of irradiated area to the redistribution area. In the present case, as the irradiated area is fixed at $\pi R_p^2$, the redistribution parameter $\rm f$ should entirely depend on the area of redistribution  by an inverse relation. The three possible cases of heat redistribution are as follows:
\begin{enumerate}
\item When the irradiated energy is redistributed over the whole planetary surface, then $\rm f=\frac{\pi R_p^2}{4\pi R_p^2} = \frac{1}{4}$ as mentioned in \cite{parmentier2015non}

\item For only dayside heat redistribution, $\rm f=\frac{\pi R_p^2}{2\pi R_p^2} = \frac{1}{2}$

\item For no heat redistribution at all, \citep{burrows2008theoretical,seager2010exoplanet} suggests  $\rm f=\frac{2}{3}$
\end{enumerate}

It is worth noting that the value of $\rm f$ decreases with the increase in the heat redistribution from the dayside to the nightside. Thus from equation \eqref{full secondary eclipse flux 2} it is evident that the re-radiated flux at the dayside increases with the decrease in the redistribution of the irradiated energy. 

\section{Numerical methodology}\label{methodology}

\subsection{Atmospheric temperature-pressure profiles}
The variation of temperature as well as pressure with atmospheric height is considered an involved and unique character of any exoplanetary atmosphere. This temperature-pressure (T-P) profiles can be affected by internal energy of the planet, irradiated flux from its host star, atmospheric redistribution, molecular mixing ratio etc. For a {hot Jupiter} planet, these (T-P) profiles are well studied by \cite{guillot2010radiative}, \cite{parmentier2014non},  \cite{parmentier2015non} in the presence of internal as well as irradiated flux on the planetary non-grey atmosphere. We use the analytical expressions prescribed by them in order to study the effect of redistribution on planetary atmosphere. The study of the variation of the gas mixing ratio on the atmospheric redistribution is however beyond the scope of the present work.

The atmospheric temperature-pressure profiles are generated using the numerical code developed by \cite{parmentier2015non} and available in public domain\footnote{\href{http://cdsarc.u-strasbg.fr/viz-bin/qcat?J/A+A/574/A35}{http://cdsarc.u-strasbg.fr/viz-bin/qcat?J/A+A/574/A35}}. The numerical code is based on the analytical model given in \cite{guillot2010radiative, parmentier2014non} and can be expressed by equation \eqref{TP for collimated}. We have ignored the convective region at the bottom of the model T-P profiles. The derived T-P profiles are used to solve the radiative tranfer equations numerically in order to obtain the planetary emission spectra. Since the emission spectra probe  deep of the atmosphere, therefore, unlike the transmission and reflection spectra, the T-P profile up to a deeper atmospheric region is required to calculate the emission flux. Hence, we consider the T-P profile for a wide range  of 
{atmospheric}
pressure, e.g.,  $10^{-6}-10^2$ bar. Also the profiles are generated for a Jupiter like planet with surface gravity $25 m/s^2$ and  internal temperature 200K but equilibrium temperature of 1800K for zero albedo. The opacity is considered to be Rosseland mean opacity as provided by \cite{valencia2013bulk}. Finally all the temperature-pressure profiles are calculated in presence of {the solar composition} VO and TiO so that the effect of thermal inversion could be incorporated. These T-P profiles are used to generate the absorption and scattering co-efficients as well as to solve the radiative transfer equations.

\subsection{ Absorption and scattering co-efficients}
In the present work we aim to study the influence of heat redistribution on the atmosphere of  {hot Jupiter}s. Now an exoplanetary atmosphere can be characterized by its temperature-pressure profile, atmospheric chemistry and atmospheric emission. Here we investigate how the temperature-pressure profile as well as atmospheric emission varies with the atmospheric heat redistribution for a fixed atmospheric chemistry. Hence, we present models with fixed abundance, e.g.,  solar abundance of atoms and molecules for {hot Jupiter}s. The absorption as well as scattering co-efficients are calculated using the numerical package "Exo-Transmit" developed by \cite{kempton2017exo} and available in public domain\footnote{\href{https://github.com/elizakempton/Exo_Transmit}{https://github.com/elizakempton/Exo\_Transmit}}  along with the atomic and molecular database provided in \cite{freedman2008line, freedman2014gaseous, lupu2014atmospheres}. The model description as well as validity check of "Exo-Transmit" software package with Tau-REx package  \citep{waldmann2015tau} is discussed in \cite{sengupta2020optical}. 
We fix the chemical composition by choosing solar abundance equation of state (EOS) data provided with the package. For this particular chemical composition, the absorption and scattering co-efficients are generated for different temperature-pressure profiles. These co-efficients are further used to calculate the single scattering albedo at each temperature-pressure point for different wavelengths. We ignore cloud or condensate opacities or any collision induced scattering.


\subsection{Numerical method to generate the emission spectra:}
In order to generate the synthetic emission spectra, we need to solve the radiative transfer equations by using the atmospheric temperature-pressure profile as well as the absorption and scattering co-efficients. The radiative transfer equations appropriate for an atmosphere in thermodynamic equilibrium can be expressed as,
\begin{equation}\label{radiative transfer}
\mu\frac{dI(\mu,\nu,\tau)}{d\tau} = I(\mu,\nu,\tau) - \xi(\mu,\tau,\nu)
\end{equation}
where, $\xi(\mu,\tau,\nu)$ is the source function. We consider local thermodynamic equilibrium at each atmospheric layer and thus the source function can be written as $\xi(\mu,\tau,\nu) = B(\nu,T) = \frac{2h\nu^3}{c^2} (e^{h\nu/kT}-1)^{-1}$ \citep{Chandrasekhar, seager2010exoplanet}. Thus, each atmospheric layer contributes to the specific intensity $I$ in terms of their blackbody temperature. To solve this equations numerically we use the formalism of discrete space theory as developed by \cite{peraiah1973numerical},\cite{sengupta2009multiple}.
The scalar version of the numerical code {is used} through the following steps,
\begin{enumerate}
    \item The atmosphere of the planet is divided into many ”shells”, each {of} them is in Local Thermodynamic Equilibrium {(LTE)} and their thickness is less than or equal to a critical thickness $\tau_{critical}$ which is calculated on the basis of the physical characteristics of the medium. {In this work the number of atmospheric shells vary within the range of 75 to 100. It is worth noting that, this number is decided depending on the atmospheric pressure range considered for simulation and the rate of change of the vertical temperature so that LTE would be valid at each shell.}
    \item The integration of the transfer equation is performed on each ”shell” which is a 2-dimensional grid bounded by $[r_n , r_{n+1} ]$ x $[\mu_{j - 1/2} , \mu_{j+1/2} ]$ , where, $r_n$ is the radial grid and $\mu_{j+1/2}$ is the angular grid.
    $$\mu_{j+1/2} = \sum_{k=1}^j c_k , j = 1,2,....,J$$
    where, $c_k$ are the weights of Gauss Legendre quadrature formula.
    \item {As there is an interaction between the radiation and the atmospheric shells, so the emergent radiation from a particular shell can be represented in terms of the incident radiation and the reflection and the transmission operators of the same shell. This is commonly known as the "interaction principle". We use the same mathematical formulation of this principle as given in \citep{peraiah2002introduction,sengupta2009multiple}}.
    \item {Next} we compare the discrete equations with the canonical equations of the interaction principle and found the transmission and reflection operators of the shell. {We define a matrix that contains the reflection and transmission operators of a particular shell, called as "CELL"-matrix for that particular shell \citep{peraiah2002introduction}.}
    \item {After deriving the CELL matrices for different atmospheric shells we combine them to obtain the final radiation. To combine the CELL matrices, we use the "STAR" algorithm as described in \citep{peraiah1973numerical,peraiah2002introduction,sengupta2009multiple}}.
\end{enumerate}
The runtime and efficiency of this code is provided in \cite{sengupta2020optical} in a greater detail.

\subsection{Validation of our model}\label{model validation}
\begin{figure}
\begin{center}
\begin{subfigure}[b]{0.4\textwidth}
		\centering
		\includegraphics[width=7cm,height= 4cm]{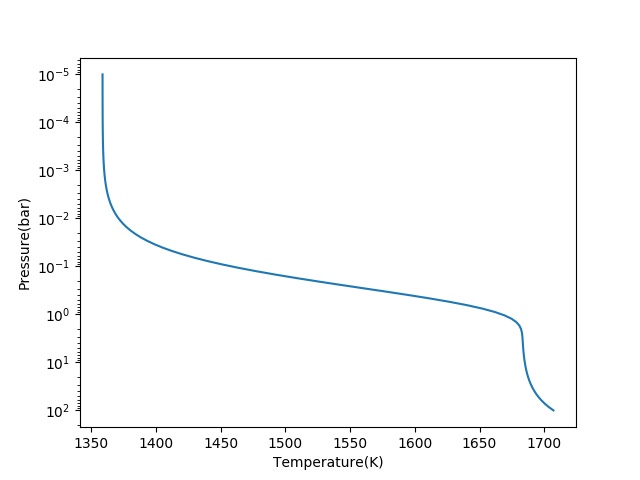}
		\caption{Without inversion}
		\label{Decreasing}
\end{subfigure}
\hfill
\begin{subfigure}[b]{0.4\textwidth}
		\includegraphics[width = 7cm,height= 4cm]{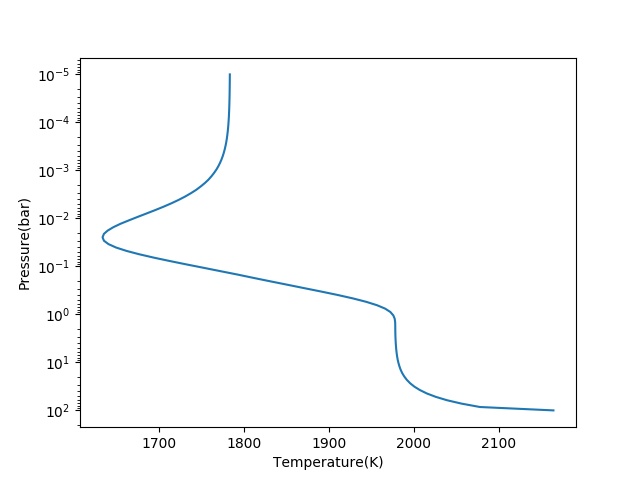}
		\caption{With inversion}
		\label{Inversion}
	\end{subfigure}
\vfill
\begin{subfigure}[b]{0.3\textwidth}
\includegraphics[width=7 cm,height= 4cm]{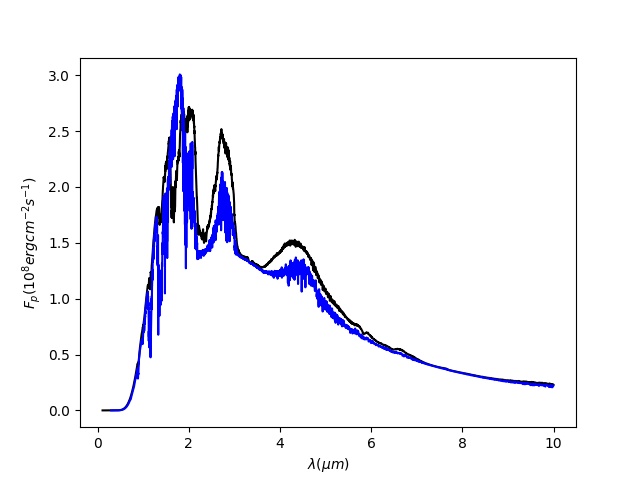}
\caption{Emission spectra}
\label{decrasing spectra_plot}
\end{subfigure}
\hfill
\begin{subfigure}[b]{0.4\textwidth}
\includegraphics[width=7 cm,height= 4cm]{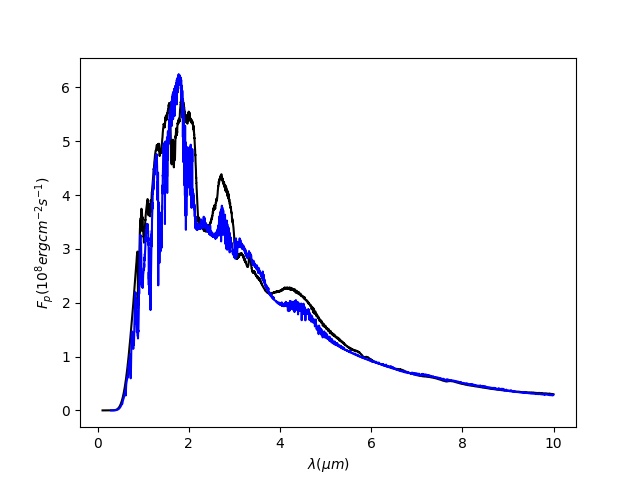}
\caption{Emission Spectra}
\label{inversion spectra_plot}
\end{subfigure}
\vfill
\begin{subfigure}[b]{0.3\textwidth}
\includegraphics[width=8 cm]{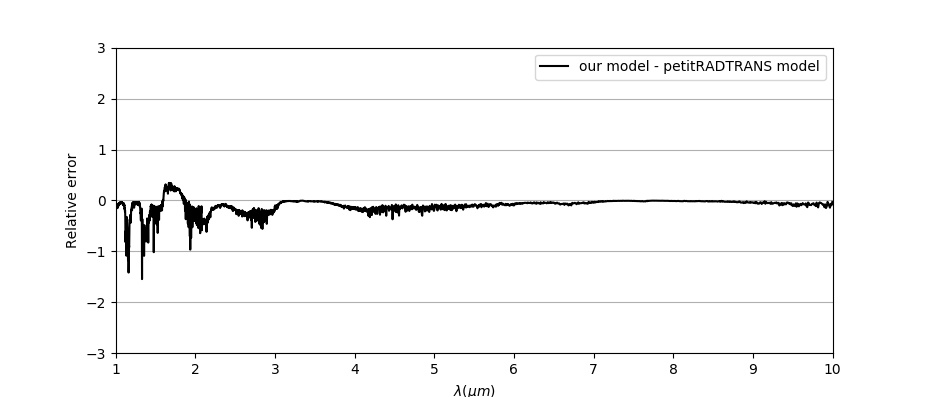}
\caption{Relative error}
\label{decreasing deviation}
\end{subfigure}
\hfill
\begin{subfigure}[b]{0.4\textwidth}
\includegraphics[width=8 cm]{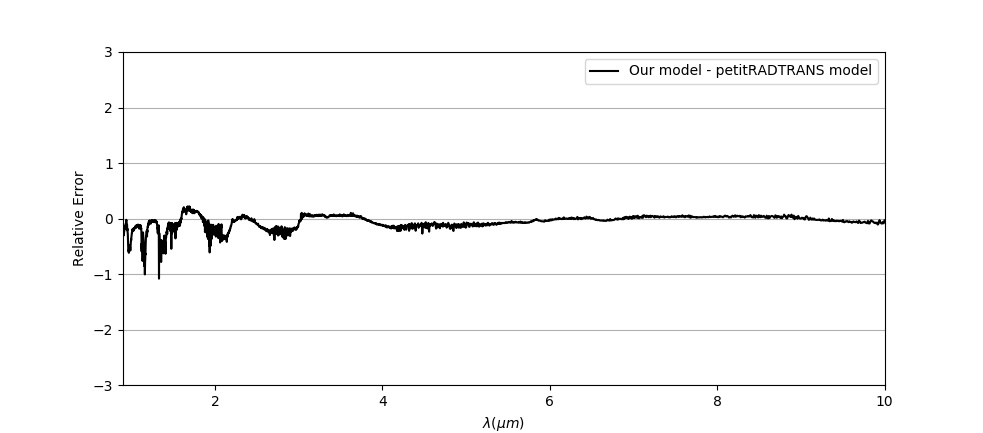}
\caption{Relative error}
\label{inversion deviation}
\end{subfigure}

\caption{Comparison of our model emission spectra (blue curve) with the model spectra calculated by using petitRADTRANS code (black curve) \citep{molliere2019petitradtrans}  without thermal inversion ( $ T_{int}=200K$ and $T_{eq} = 1500K$; left panel) and with thermal inversion ($T_{int}=200K$ and $T_{eq}=1800K$; right panel).
In uppermost panel the temperature-pressure profiles are presented.  In the middle panel (\ref{decrasing spectra_plot} and \ref{inversion spectra_plot}) the emission spectra are compared. The relative error of these two models are presented in \ref{decreasing deviation} and \ref{inversion deviation}. }
\end{center}
\end{figure}

In order to validate our numerical derivations,  we compare the emission spectra with the synthetic spectra generated by the numerical package developed by \cite{molliere2019petitradtrans}  and available in the public domain\footnote{ \href{https://gitlab.com/mauricemolli/petitRADTRANS}{https://gitlab.com/mauricemolli/petitRADTRANS}}. This software package is itself benchmarked with the petitCODE \cite{molliere2015model}. The comparisons of the two different model spectra are given in figure \ref{decrasing spectra_plot} and figure \ref{inversion spectra_plot}. In our model calculations we consider a planetary atmosphere with abundance  9.72 x $10^{-1}$, 2.3 x $10^{-2}$,  6.3 x $10^{-4}$ and 2.9x $10^{-5}$ for the elements He, $\rm H_2$, $\rm CH_4$, $NH_3$ respectively, with surface gravity
${25m/s^{-2}}$.
The temperature-pressure profiles with and without thermal inversion are  presented in figure \ref{Decreasing} and figure \ref{Inversion} respectively. We note that, in figure \ref{decreasing deviation} and \ref{inversion deviation} the relative error between these two model spectra are very small implying good agreement.  The small mismatch in the emission spectra presented in figure \ref{decrasing spectra_plot} and figure \ref{inversion spectra_plot} is due to the fact that we solved the radiative transfer equations using line by line method whereas \cite{molliere2019petitradtrans} solved the same using correlated-k approximation {(for a full phase comparison between these two methods see \cite{molliere2019petitradtrans})}.
 


\begin{figure}
\begin{center}
\begin{subfigure}[b]{0.8\textwidth}
\includegraphics[width=13cm, height=8cm]{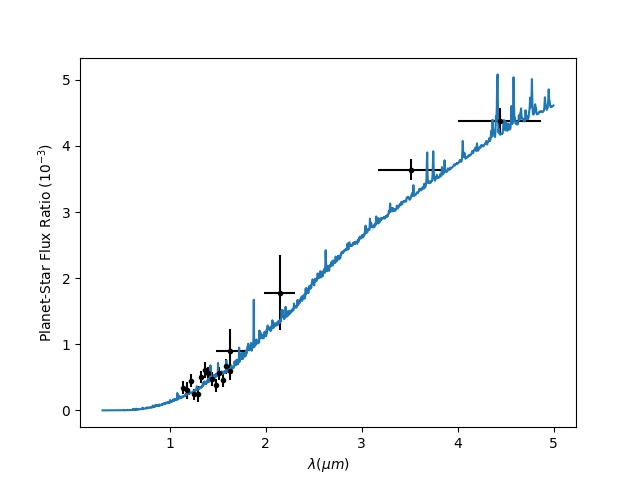}
\caption{HAT-P-32b Flux Ratio at Secondary Eclipse}
\label{HAT-P-32b secondary eclipse}
\end{subfigure}
\vfill
\begin{subfigure}[b]{0.8\textwidth}
\includegraphics[width=13cm, height=8cm]{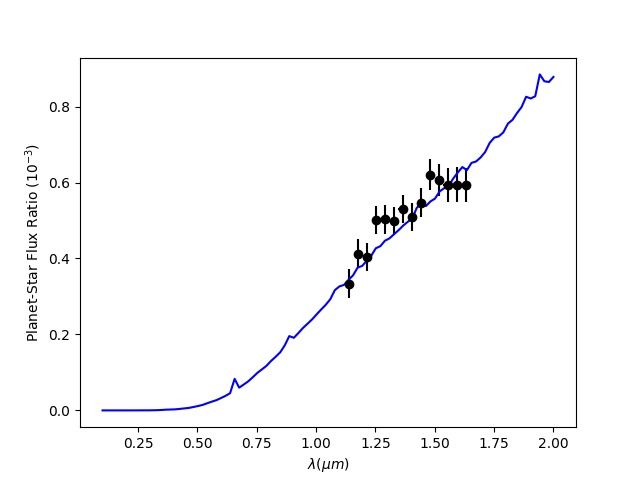}
\caption{HAT-P-7b Flux Ratio at Secondary Eclipse}
\label{HAT-P-7b secondary eclipse}
\end{subfigure}
\end{center}
\caption{Comparison of our modeled planet-to-star flux ratio (in blue) with the  observed HST/WFC3 spectrum (in black). In the upper panel \ref{HAT-P-32b secondary eclipse} the planet-to-star flux ratio for HAT-P-32b  is presented by considering an isothermal atmosphere with $T_p=1995$K. The observed data for HAT-P-32b is taken from \cite{nikolov2018hubble}.
In the lower panel \ref{HAT-P-7b secondary eclipse}, the planet-to-star flux ratio for  HAT-P-7b  is presented by considering an isothermal planetary atmosphere with $T_p=2692K$.  The observed data for HAT-P-7b is presented by \cite{mansfield2018hst}. In both the cases, solar composition is adopted.}
\end{figure}

Next we compare our model emission spectra with the observed planet-to-star flux ratio for the {hot Jupiter}s HAT-P-32b \citep{nikolov2018hubble}and HAT-P-7b  \citep{mansfield2018hst}. The planet-star flux ratio is calculated as a function of wavelength $\lambda$ from the relationship \citep{tinetti2013spectroscopy},
\begin{equation} \label{eclipse depth}
\eta(\lambda) = \frac{F_p(\lambda)}{F_s(\lambda)} \frac{R^2_p}{R^2_s}
\end{equation}
 Here, $R_p$ and $R_s $ are the  planetary and stellar radius respectively. $F_p$ is the same as $F_{full}$ in equation \eqref{full secondary eclipse flux 2}. Thus the emission spectra is dependent on the redistribution parameter $\rm f$.
 While calculating the emission spectra by using equation \eqref{eclipse depth}, we have used the stellar flux $F_s$ of 
 {PHOENIX}
  models \citep{husser2013new}. For HAT-P-32b, the planetary surface gravity is taken to be 
$g=6.6 {m s^{-2}}$
\citep{hartman2011hat}, and an isothermal atmosphere with $T_{p}$=1995K \citep{nikolov2018hubble} is considered. $R_p/R_{star}$ is fixed at = 0.1506. The result is presented in figure \ref{HAT-P-32b secondary eclipse}.  For the case of HAT-P-7b, an isothermal atmosphere with $T_{p}$=2692K \citep{mansfield2018hst} is considered with the surface gravity 
${g=20 m s^{-2}}$
\citep{stassun2017accurate} and $\frac{R_p}{R_{star}} = 0.07809$	\citep{wong20163}. The observed data for the planet-star flux ratio in this case was obtained using HST/WFC3 camera within the wavelength range 1.1-1.7 $\mu m$ \citep{mansfield2018hst}. The model spectrum along with the observed data is presented in figure \ref{HAT-P-7b secondary eclipse}. For both the cases, our model emission spectra fits reasonably well with the observed data within the 
{error bars}.
 

\section{Results}\label{Results}

\subsection{Effect of redistribution parameter on T-P profiles and on emission spectra}\label{Effect of f on TP and spectra}
We investigate the effect of atmospheric heat redistribution on the T-P profiles as well as on the emission spectra during the secondary eclipse of {hot Jupiter}s by using the procedure described in section~\ref{methodology}. Without loss of generality, we fixed the internal temperature $T_{int}=200K$ and the equilibrium temperature for zero albedo $T_{eq0}=1800K$, surface gravity $g=25 m/s^2$. We use Rosseland mean opacity given in \cite{valencia2013bulk} and included TiO and VO {as present in the solar composition \cite{parmentier2015non}}. The Bond albedo is  fixed for different redistribution cases and its values are calculated from the fit given in \cite{parmentier2014non}.  We have not considered the convective solution as provided in \cite{parmentier2015non} and only the radiative solution is considered. 

\begin{figure}
\begin{center}
\begin{subfigure}[b]{\textwidth}
\includegraphics[width=14cm,height=8cm]{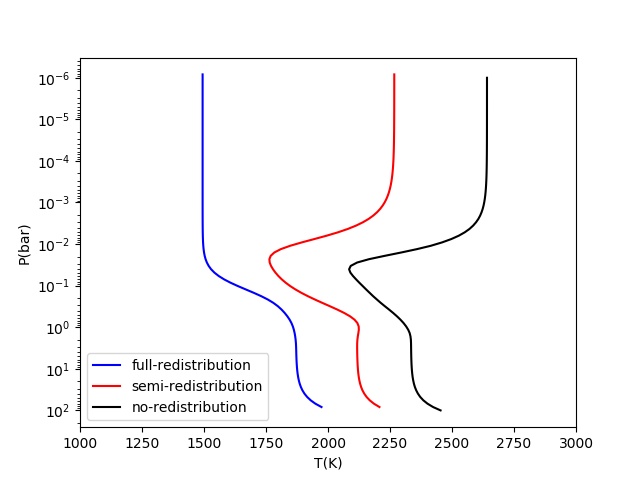}
\caption{Temperature-pressure profile}
\label{redistribution_TP}
\end{subfigure}
\vfill
\begin{subfigure}[b]{\textwidth}
\includegraphics[width=14cm,height=8cm]{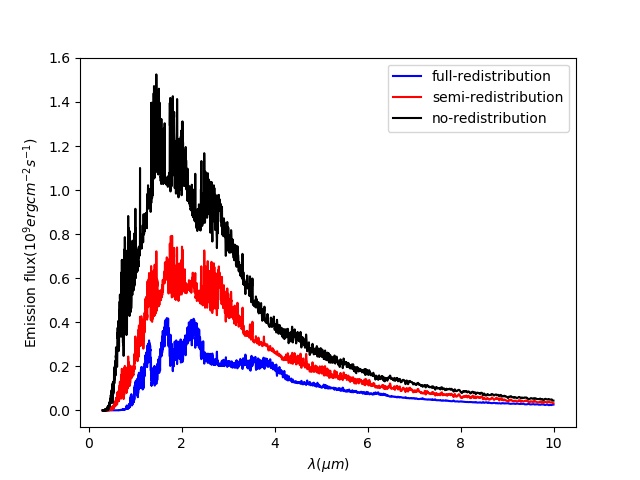}
\caption{Emission spectra}
\label{redistribution_emsn}
\end{subfigure}
\caption{Effect of heat redistribution on the T-P profiles and on the planetary emission spectra. The upper panel shows the T-P profiles for different values of the redistribution parameter $\rm f=0.25$ (full-redistribution), $\rm f=0.5$ (semi-redistribution) and $\rm f=2/3$ (no-redistribution). In the lower panel, the emission spectra during the secondary eclipse is presented with different atmospheric redistribution.}
\end{center}
\end{figure}

\begin{figure}[ht!]
\begin{center}
\includegraphics[scale=0.7]{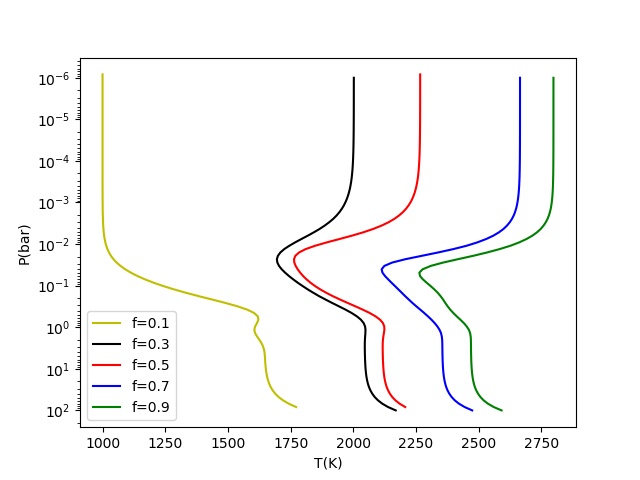}
\caption{T-P profiles for different values of the heat redistribution parameter $\rm f$. The planetary parameters adopted are (i) equilibrium temperature $T_{eq}=1800K$ (ii) internal temperature $T_{int}=200K$ and surface gravity $g=25$ $\rm ms^{-2}$. Solar composition with TiO is assumed}
\label{redistribution dependent TP}
\end{center}
\end{figure}

 First we consider the three different heat redistribution cases under isotropic approximation as discussed in section~\ref{f for isotropic approximation}. In figure \ref{redistribution_TP} we present the T-P profiles under these three special conditions, e.g., full, semi and no heat redistribution. The corresponding emission flux for an atmosphere with solar abundances are presented in figure  \ref{redistribution_emsn} with the same conditions as mentioned earlier. 
Next we consider a general case of simulation for the redistribution parameter $\rm f$ varying from 0.1 to 0.9 with a constant interval of 0.2. The corresponding T-P profiles are given in figure \ref{redistribution dependent TP}. These are calculated by using the formalism provided in \cite{parmentier2015non}.

\begin{figure}[ht!]
\begin{center}
\includegraphics[scale=0.7]{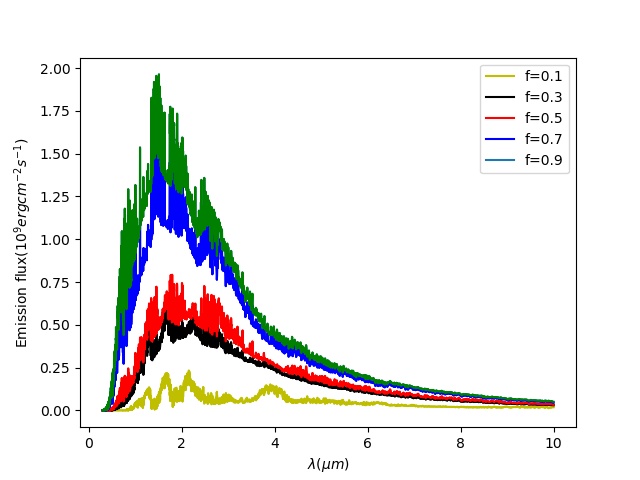}
\caption{ Planetary emission spectra for different values of atmospheric heat redistribution parameter $\rm f$.}
\label{redistribution dependent emsn}
\end{center}
\end{figure}

 According to equation \eqref{full secondary eclipse flux 2}  the total flux emitted from the substellar point of the planet at full phase is directly proportional to the heat redistribution parameter $\rm f$. With the change in the T-P profiles, the emission spectra alter with different values of the heat redistribution parameter $\rm f$. This variation is studied by solving the radiative transfer equations numerically for an atmosphere with solar abundances and presented in figure \ref{redistribution dependent emsn}.

\subsection{Case study: Emission Spectra of exoplanet XO-1b:}\label{XO-1b case}
Finally we model the emission spectra of the {hot Jupiter} XO-1b with atmospheric heat redistribution. Using IRAC of Spitzer Telescope, \cite{machalek2008thermal} for the first time observed this planet during secondary eclipse.
{The observation was done at wavelength points 3.6, 4.5, 5.8, and 8.0 $\mu m$ which is within the wavelength range of our modeled spectra (i.e. 0 - 10 $\mu m$). Also the temperature, mass and radius of the host star XO-1 is similar to that of the sun. This similarity gives the opportunity for accurate modeling as the PHOENIX model \cite{husser2013new}, which we used here for generating the stellar spectra, works best at the solar parameters. For these reasons we choose XO-1b as a test case in our study.}

{Subsequently,}
\cite{tinetti2010probing} obtained the NIR transmission spectra of XO-1b by probing the terminator region. In their study,\cite{tinetti2010probing} considered  the planet mass $M_p=0.9\pm0.07 M_J$, planet-star radius ratio $R_p/R_{star} =0.1326 \pm 0.0004$, star-planet distance $a= 0.04928\pm 0.00089$ AU, effective temperature of the host star $T_{eff} = 5750K$ and planetary equilibrium temperature $T_{eq}=1200K$. From transmission spectra analysis, they estimated the best fitted atmospheric abundances to be $H_2O\approx4.5 X 10^{-4}$, $CH_4 \approx 10^{-5}$, $CO_2 \approx 4.5 X 10^{-4}$ and $CO\approx 10^{-2}$.
However, while retrieving the T-P profile, they found a degeneracy that T-P profiles for both thermal inversion and non-inversion could explain the transmission spectra accurately (see  figure 3 of \cite{tinetti2010probing}).  Since the transmission spectra is not very sensitive to the atmospheric T-P profile \citep{sengupta2020optical}, the dayside emission spectra may serve as  a potential tool to remove this degeneracy. \cite{tinetti2010probing} calculated the emission spectra of the planet by assuming total atmospheric heat redistribution i.e., by keeping the chemical composition and T-P profile the same in the terminator region as well as in the dayside of the planet. We adopt the same condition here and calculated the dayside emission spectra by solving line by line radiative transfer equations using the discrete space theory formalism.

\begin{figure}
\begin{center}
\begin{subfigure}[b]{0.7\textwidth}
\includegraphics[width=9cm]{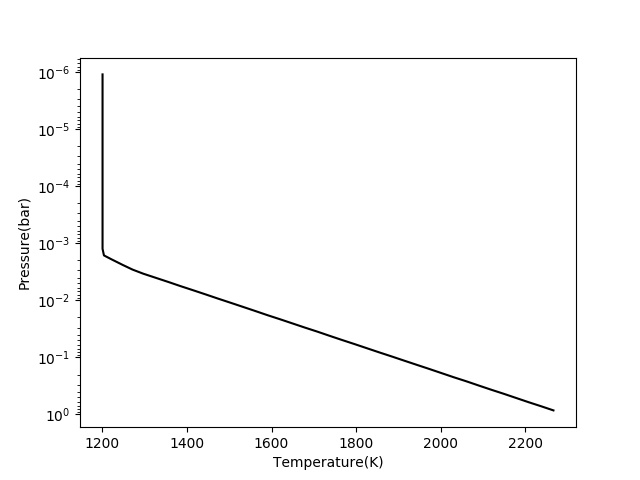}
\caption{T-P profile}
\label{XO-1b No inversion TP}
\end{subfigure}
\vfill
\begin{subfigure}[b]{1.4\textwidth}
\includegraphics[width=16cm]{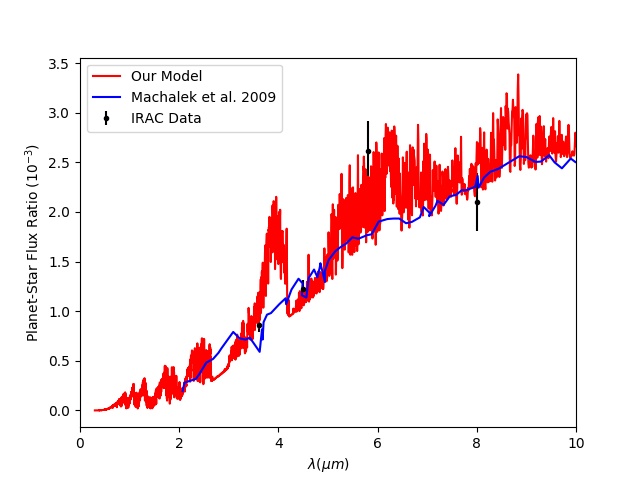}
\caption{Flux Ratio at Secondary Eclipse}
\label{XO-1b_secondary_eclipse_no_inversion}
\end{subfigure}
\caption{Left Panel: The T-P profile of {hot Jupiter} XO-1b with no thermal inversion appropriate for the night side atmospheric composition derived by \cite{tinetti2010probing}. Right Panel: A comparison of our modeled emission spectrum {(in red)} corresponding to this T-P profile (see left panel) and the atmospheric composition derived in \cite{tinetti2010probing}, the observed IRAC emission flux and the model emission spectrum {(in blue)} presented by \cite{machalek2008thermal}.}
\end{center}
\end{figure}

We have calculated the EOS by using the atmospheric mixing ratios as described above and used them in the Exo-Transmit package to estimate the absorption and scattering co-efficients. In the next step, we solve the radiative transfer equations using the T-P profile presented in figure \ref{XO-1b No inversion TP} and calculate the emission spectrum for XO-1b. Finally, by using PHOENIX model spectrum \cite{husser2013new} of the star with effective temperature 5750K, we calculated the planet-to-star flux ratio. In figure \ref{XO-1b_secondary_eclipse_no_inversion} we present our model emission spectrum along with the observed data of IRAC and the model spectrum presented by \cite{machalek2008thermal}.

\section{Discussions \& Conclusion}\label{discussion}

We demonstrate the effect of atmospheric heat redistribution on the thermal properties of {hot Jupiter}'s e.g., on the temperature distribution at different height of the atmosphere and the dayside emission spectra.  
For the simplest case of isotropic approximation of the incident flux, the heat redistribution parameter $\rm f$ can have three different values depending on the degree of heat redistribution. The value of the redistribution parameter $\rm f$ decreases as the amount of heat redistribution increases. When heat redistribution is less, the temperature of the dayside atmosphere  is higher because less amount of heat is transfered from sub-stellar side to anti-stellar side. Fig.~\ref{redistribution_TP} shows that when the heat redistribution reduces, the entire T-P profile shifts towards a higher temperature.  For a full redistribution ($\rm f=1/4$), the vertical temperature profile    does not show any thermal inversion whereas for semi-redistribution ($f=1/2$) and no redistribution ($f=2/3$), the T-P profile shows significant thermal inversion in the upper atmospheric region.  Similarly, figure \ref{redistribution_emsn} shows that the emission flux  is maximum when there is  no heat redistribution and minimum when the heat is fully redistributed. This is because of the fact that the less the heat redistributed, the less amount of heat is transferred from dayside to nightside and hence greater  amount of radiation emerges out from the dayside. For a fixed atmospheric composition, the features of the emission spectra remains unaltered. 

For isotropic approximation, only three cases of heat redistribution may take place. But for more realistic situation, the amount of heat redistribution parametrized by the  redistribution parameter $\rm f$ can take any values within the range $0\leq f \leq 1$. So in the next part of our work we have 
{simulated}
the whole simulation for the redistribution parameter values of 0.1, 0.3, 0.5, 0.7 and 0.9. The corresponding T-P profiles as well as the emission profiles are shown in fig.~\ref{redistribution dependent TP} and \ref{redistribution dependent emsn} respectively.  While comparing figures \ref{redistribution_TP} and \ref{redistribution dependent TP}, we notice that the T-P profiles start showing thermal inversion when $\rm f>0.25$. The temperature increases significantly with the increase in the value of $\rm f$  i.e.,  with the decrease in heat redistribution. Consequently, the emission flux increases with the increase in the redistribution parameter $\rm f $ i.e., with the decrease in heat redistribution as shown in figure~\ref{redistribution dependent emsn}. This behaviour of the emission flux is obvious  from equation~\eqref{full secondary eclipse flux 2}. 
Since, the planet-to-star flux ratio increases with the increase in the value of $\rm f$,  i.e. with the decrease in the heat redistribution, a model fit of the observed dayside emission spectra may provide good idea on the amount of heat redistribution in the planetary atmosphere.

Finally, we apply our analysis to the exoplanet XO-1b. The transmission spectra of this planet was observed by \cite{tinetti2010probing} at the terminator region and the T-P profile is retrieved thereby  as shown in figure~\ref{XO-1b No inversion TP}. The absence of thermal inversion in the retrieved T-P profile implies almost full atmospheric heat redistribution. We compare the observed planet-to-star flux ratio with our model spectrum by using the same T-P profile. Figure~\ref{XO-1b_secondary_eclipse_no_inversion} shows that our modeled planet-to-star flux ratio matches with the observed data better than that of \cite{machalek2008thermal}. {Hence} it can be inferred that almost full heat redistribution takes place in the atmosphere of XO-1b.

 In this work we studied the effect of heat redistribution from sub-stellar side to anti-stellar side on the T-P profile and on the planetary emission spectra of hot gas giant planets. However, according to the chemical equilibrium, the mixing ratios of various atomic and molecular species may differ from the dayside to nightside of the atmosphere due to the difference in temperature owing to insignificant heat redistribution.
{In that case the Collision Induced Absorption (CIA) can also be included for a more realistic approach as the amount of CIA will change with the amount of heat redistribution and introduce additional features in the emission spectra.}
 For instance, \cite{tan2019atmospheric} considered abundance of atomic hydrogen in the dayside of the atmosphere and molecular hydrogen in the nightside of a hydrogen rich hot gas giant planet to describe the mechanism of heat redistribution. Therefore, such compositional difference may be important to understand the amount of heat redistribution from the observed emission spectra. 

In this study, we have considered the LTE condition while solving the radiative transfer equation~\eqref{radiative transfer} and thus the source term $\xi$ becomes the Planck emission which is only a function of the temperature of the layer \cite{Chandrasekhar}. But the emitted  radiation from one layer would be further scattered by the other atmospheric layers as well. However  \cite{goldstein1960infrared} showed that infrared reflectivity is an important phenomena in case of Rayleigh scattering of planetary atmosphere. Thus, the scattering term should be included along with the thermal emission in the transfer equation as shown in \cite{bellman1967chandrasekhar}, \cite{sengupta2021effects},\cite{sengupta2022atmospheric} while modeling the emission spectra.

{It is indeed the first step towards studying the atmospheric heat redistribution in terms of the day side emission spectra of hot Jupiters. But the heat redistribution will also have some impact on the reflection and transmission spectra. Hence the redistribution can also be studied in terms of those spectra.}

The effect of atmospheric heat redistribution on the emission curve during secondary eclipse was not detectable by the early observation facilities (e.g. Spitzer). However,\cite{komacek2019temporal} argued that the secondary eclipse depth variability $\le 2\%$ can be detected using the
{recently launched JWST and }
future telescope like ARIEL. 
{Note that, it is out of scope of the present theoretical study to explore the observational requirements for detecting the variation of dayside emission spectra due to the horizontal atmospheric heat redistribution. Hence an explicit study on present and future observational facilities compatible for this kind of detection are needed to predict the amount of atmospheric heat redistribution from the emission spectra observations.}
It is expected that the detections of variation of secondary eclipse emission spectra due to atmospheric {heat} redistribution will be possible using these telescopes in near future. 


{The detection of minute changes in the emission spectra caused by the atmospheric heat redistribution is a big challenge in observational astronomy. But the estimation of observational parameters is very important for a quantitative study of the planetary atmosphere. Hence, further studies in this direction along with the telescopic parameters are needed to give proper targets for present and future telescopes.}

{Softwares used: Exo\_Transmit \citep{kempton2017exo,freedman2008line,freedman2014gaseous,lupu2014atmospheres}, simpy \citep{meurer2017sympy}, petitRADTRANS \citep{molliere2019petitradtrans},  Analytical model for irradiated atmosphere \citep{parmentier2015non,parmentier2014non,valencia2013bulk,guillot2010radiative}}

\bibliography{paper2}

\begin{thebibliography}{10}
\expandafter\ifx\csname url\endcsname\relax
  \def\url#1{\texttt{#1}}\fi
\expandafter\ifx\csname urlprefix\endcsname\relax\def\urlprefix{URL }\fi
\expandafter\ifx\csname href\endcsname\relax
  \def\href#1#2{#2} \def\path#1{#1}\fi

\bibitem{mayor1995jupiter}
M.~Mayor, D.~Queloz, A jupiter-mass companion to a solar-type star, Nature
  378~(6555) (1995) 355--359.

\bibitem{mazeh2000spectroscopic}
T.~Mazeh, D.~Naef, G.~Torres, D.~W. Latham, M.~Mayor, J.-L. Beuzit, T.~M.
  Brown, L.~Buchhave, M.~Burnet, B.~W. Carney, et~al., The spectroscopic orbit
  of the planetary companion transiting hd 209458, The Astrophysical Journal
  Letters 532~(1) (2000) L55.

\bibitem{knutson2007map}
H.~A. Knutson, D.~Charbonneau, L.~E. Allen, J.~J. Fortney, E.~Agol, N.~B.
  Cowan, A.~P. Showman, C.~S. Cooper, S.~T. Megeath, A map of the day--night
  contrast of the extrasolar planet hd 189733b, Nature 447~(7141) (2007)
  183--186.

\bibitem{hansen2008absorption}
B.~M. Hansen, On the absorption and redistribution of energy in irradiated
  planets, The Astrophysical Journal Supplement Series 179~(2) (2008) 484.

\bibitem{guillot2010radiative}
T.~Guillot, On the radiative equilibrium of irradiated planetary atmospheres,
  Astronomy \& Astrophysics 520 (2010) A27.

\bibitem{parmentier2014non}
V.~Parmentier, T.~Guillot, A non-grey analytical model for irradiated
  atmospheres-i. derivation, Astronomy \& Astrophysics 562 (2014) A133.

\bibitem{parmentier2015non}
V.~Parmentier, T.~Guillot, J.~J. Fortney, M.~S. Marley, A non-grey analytical
  model for irradiated atmospheres-ii. analytical vs. numerical solutions,
  Astronomy \& Astrophysics 574 (2015) A35.

\bibitem{burrows2008optical}
A.~Burrows, L.~Ibgui, I.~Hubeny, Optical albedo theory of strongly irradiated
  giant planets: The case of hd 209458b, The Astrophysical Journal 682~(2)
  (2008) 1277.

\bibitem{tinetti2013spectroscopy}
G.~Tinetti, T.~Encrenaz, A.~Coustenis, Spectroscopy of planetary atmospheres in
  our galaxy, The Astronomy and Astrophysics Review 21~(1) (2013) 1--65.

\bibitem{molliere2019petitradtrans}
P.~Molli{\`e}re, J.~Wardenier, R.~van Boekel, T.~Henning, K.~Molaverdikhani,
  I.~Snellen, petitradtrans-a python radiative transfer package for exoplanet
  characterization and retrieval, Astronomy \& Astrophysics 627 (2019) A67.

\bibitem{malik2017helios}
M.~Malik, L.~Grosheintz, J.~M. Mendon{\c{c}}a, S.~L. Grimm, B.~Lavie,
  D.~Kitzmann, S.-M. Tsai, A.~Burrows, L.~Kreidberg, M.~Bedell, et~al., Helios:
  an open-source, gpu-accelerated radiative transfer code for self-consistent
  exoplanetary atmospheres, The astronomical journal 153~(2) (2017) 56.

\bibitem{drummond2018effect}
B.~Drummond, N.~Mayne, I.~Baraffe, P.~Tremblin, J.~Manners, D.~S. Amundsen,
  J.~Goyal, D.~Acreman, The effect of metallicity on the atmospheres of
  exoplanets with fully coupled 3d hydrodynamics, equilibrium chemistry, and
  radiative transfer, Astronomy \& Astrophysics 612 (2018) A105.

\bibitem{gandhi2017genesis}
S.~Gandhi, N.~Madhusudhan, Genesis: new self-consistent models of exoplanetary
  spectra, Monthly Notices of the Royal Astronomical Society 472~(2) (2017)
  2334--2355.

\bibitem{sengupta2009multiple}
S.~Sengupta, M.~S. Marley, Multiple scattering polarization of substellar-mass
  objects: T dwarfs, The Astrophysical Journal 707~(1) (2009) 716.

\bibitem{tinetti2007water}
G.~Tinetti, A.~Vidal-Madjar, M.-C. Liang, J.-P. Beaulieu, Y.~Yung, S.~Carey,
  R.~J. Barber, J.~Tennyson, I.~Ribas, N.~Allard, et~al., Water vapour in the
  atmosphere of a transiting extrasolar planet, Nature 448~(7150) (2007)
  169--171.

\bibitem{swain2008presence}
M.~R. Swain, G.~Vasisht, G.~Tinetti, The presence of methane in the atmosphere
  of an extrasolar planet, Nature 452~(7185) (2008) 329--331.

\bibitem{madhusudhan2014h2o}
N.~Madhusudhan, N.~Crouzet, P.~R. McCullough, D.~Deming, C.~Hedges, H2o
  abundances in the atmospheres of three hot jupiters, The Astrophysical
  Journal Letters 791~(1) (2014) L9.

\bibitem{fraine2014water}
J.~Fraine, D.~Deming, B.~Benneke, H.~Knutson, A.~Jord{\'a}n, N.~Espinoza,
  N.~Madhusudhan, A.~Wilkins, K.~Todorov, Water vapour absorption in the clear
  atmosphere of a neptune-sized exoplanet, Nature 513~(7519) (2014) 526--529.

\bibitem{kreidberg2014clouds}
L.~Kreidberg, J.~L. Bean, J.-M. D{\'e}sert, B.~Benneke, D.~Deming, K.~B.
  Stevenson, S.~Seager, Z.~Berta-Thompson, A.~Seifahrt, D.~Homeier, Clouds in
  the atmosphere of the super-earth exoplanet gj 1214b, Nature 505~(7481)
  (2014) 69--72.

\bibitem{barstow2016consistent}
J.~K. Barstow, S.~Aigrain, P.~G. Irwin, D.~K. Sing, A consistent retrieval
  analysis of 10 hot jupiters observed in transmission, The Astrophysical
  Journal 834~(1) (2016) 50.

\bibitem{benneke2012atmospheric}
B.~Benneke, S.~Seager, Atmospheric retrieval for super-earths: uniquely
  constraining the atmospheric composition with transmission spectroscopy, The
  Astrophysical Journal 753~(2) (2012) 100.

\bibitem{nikolov2018hubble}
N.~Nikolov, D.~Sing, J.~Goyal, G.~Henry, H.~Wakeford, T.~Evans,
  M.~L{\'o}pez-Morales, A.~Garc{\'\i}a~Mu{\~n}oz, L.~Ben-Jaffel,
  J.~Sanz-Forcada, et~al., Hubble pancet: an isothermal day-side atmosphere for
  the bloated gas-giant hat-p-32ab, Monthly Notices of the Royal Astronomical
  Society 474~(2) (2018) 1705--1717.

\bibitem{madhusudhan2009temperature}
N.~Madhusudhan, S.~Seager, A temperature and abundance retrieval method for
  exoplanet atmospheres, The Astrophysical Journal 707~(1) (2009) 24.

\bibitem{tinetti2010probing}
G.~Tinetti, P.~Deroo, M.~Swain, C.~Griffith, G.~Vasisht, L.~Brown, C.~Burke,
  P.~McCullough, Probing the terminator region atmosphere of the hot-jupiter
  xo-1b with transmission spectroscopy, The Astrophysical Journal Letters
  712~(2) (2010) L139.

\bibitem{brogi2019retrieving}
M.~Brogi, M.~R. Line, Retrieving temperatures and abundances of exoplanet
  atmospheres with high-resolution cross-correlation spectroscopy, The
  Astronomical Journal 157~(3) (2019) 114.

\bibitem{zhang2019forward}
M.~Zhang, Y.~Chachan, E.~M.-R. Kempton, H.~A. Knutson, Forward modeling and
  retrievals with platon, a fast open-source tool, Publications of the
  Astronomical Society of the Pacific 131~(997) (2019) 034501.

\bibitem{fisher2018retrieval}
C.~Fisher, K.~Heng, Retrieval analysis of 38 wfc3 transmission spectra and
  resolution of the normalization degeneracy, Monthly Notices of the Royal
  Astronomical Society 481~(4) (2018) 4698--4727.

\bibitem{tsiaras2018population}
A.~Tsiaras, I.~Waldmann, T.~Zingales, M.~Rocchetto, G.~Morello, M.~Damiano,
  K.~Karpouzas, G.~Tinetti, L.~McKemmish, J.~Tennyson, et~al., A population
  study of gaseous exoplanets, The Astronomical Journal 155~(4) (2018) 156.

\bibitem{sengupta2020optical}
S.~Sengupta, A.~Chakrabarty, G.~Tinetti, Optical transmission spectra of hot
  jupiters: effects of scattering, The Astrophysical Journal 889~(2) (2020)
  181.

\bibitem{chakrabarty2020effects}
A.~Chakrabarty, S.~Sengupta, Effects of thermal emission on the transmission
  spectra of hot jupiters, The Astrophysical Journal 898~(1) (2020) 89.

\bibitem{waldmann2015tau}
I.~P. Waldmann, G.~Tinetti, M.~Rocchetto, E.~J. Barton, S.~N. Yurchenko,
  J.~Tennyson, Tau-rex i: A next generation retrieval code for exoplanetary
  atmospheres, The Astrophysical Journal 802~(2) (2015) 107.

\bibitem{kempton2017exo}
E.~M.-R. Kempton, R.~Lupu, A.~Owusu-Asare, P.~Slough, B.~Cale, Exo-transmit: An
  open-source code for calculating transmission spectra for exoplanet
  atmospheres of varied composition, Publications of the Astronomical Society
  of the Pacific 129~(974) (2017) 044402.

\bibitem{batalha2019exoplanet}
N.~E. Batalha, M.~S. Marley, N.~K. Lewis, J.~J. Fortney, Exoplanet
  reflected-light spectroscopy with picaso, The Astrophysical Journal 878~(1)
  (2019) 70.

\bibitem{gibson2020detection}
N.~P. Gibson, S.~Merritt, S.~K. Nugroho, P.~E. Cubillos, E.~J. de~Mooij,
  T.~Mikal-Evans, L.~Fossati, J.~Lothringer, N.~Nikolov, D.~K. Sing, et~al.,
  Detection of fe i in the atmosphere of the ultra-hot jupiter wasp-121b, and a
  new likelihood-based approach for doppler-resolved spectroscopy, Monthly
  Notices of the Royal Astronomical Society 493~(2) (2020) 2215--2228.

\bibitem{waldmann2013signals}
I.~Waldmann, On signals faint and sparse: the acica algorithm for blind
  de-trending of exoplanetary transits with low signal-to-noise, The
  Astrophysical Journal 780~(1) (2013) 23.

\bibitem{gandhi2019hydra}
S.~Gandhi, N.~Madhusudhan, G.~Hawker, A.~Piette, Hydra-h: simultaneous hybrid
  retrieval of exoplanetary emission spectra, The Astronomical Journal 158~(6)
  (2019) 228.

\bibitem{shulyak2019remote}
D.~Shulyak, M.~Rengel, A.~Reiners, U.~Seemann, F.~Yan, Remote sensing of
  exoplanetary atmospheres with ground-based high-resolution near-infrared
  spectroscopy, Astronomy \& Astrophysics 629 (2019) A109.

\bibitem{madhusudhan2011high}
N.~Madhusudhan, S.~Seager, High metallicity and non-equilibrium chemistry in
  the dayside atmosphere of hot-neptune gj 436b, The Astrophysical Journal
  729~(1) (2011) 41.

\bibitem{greene2016characterizing}
T.~P. Greene, M.~R. Line, C.~Montero, J.~J. Fortney, J.~Lustig-Yaeger,
  K.~Luther, Characterizing transiting exoplanet atmospheres with jwst, The
  Astrophysical Journal 817~(1) (2016) 17.

\bibitem{blecic2017implications}
J.~Blecic, I.~Dobbs-Dixon, T.~Greene, The implications of 3d thermal structure
  on 1d atmospheric retrieval, The Astrophysical Journal 848~(2) (2017) 127.

\bibitem{line2013systematic}
M.~R. Line, A.~S. Wolf, X.~Zhang, H.~Knutson, J.~A. Kammer, E.~Ellison,
  P.~Deroo, D.~Crisp, Y.~L. Yung, A systematic retrieval analysis of secondary
  eclipse spectra. i. a comparison of atmospheric retrieval techniques, The
  Astrophysical Journal 775~(2) (2013) 137.

\bibitem{sengupta2021effects}
S.~Sengupta, Effects of thermal emission on chandrasekhar's semi-infinite
  diffuse reflection problem, The Astrophysical Journal 911~(2) (2021) 126.

\bibitem{sengupta2022atmospheric}
S.~Sengupta, Atmospheric thermal emission effect on chandrasekhar’s finite
  atmosphere problem, The Astrophysical Journal 936~(2) (2022) 139.

\bibitem{showman2007atmospheric}
A.~P. Showman, K.~Menou, J.~Y. Cho, Atmospheric circulation of hot jupiters: a
  review of current understanding, arXiv preprint arXiv:0710.2930.

\bibitem{hammond2020equatorial}
M.~Hammond, S.-M. Tsai, R.~T. Pierrehumbert, The equatorial jet speed on
  tidally locked planets. i. terrestrial planets, The Astrophysical Journal
  901~(1) (2020) 78.

\bibitem{komacek2019temporal}
T.~D. Komacek, A.~P. Showman, Temporal variability in hot jupiter atmospheres,
  The Astrophysical Journal 888~(1) (2019) 2.

\bibitem{seager2010exoplanets}
S.~Seager, Exoplanets, University of Arizona Press In collaboration with Lunar
  and Planetary Institute, Tucson Houston, 2010.

\bibitem{heng2015atmospheric}
K.~Heng, A.~P. Showman, Atmospheric dynamics of hot exoplanets, Annual Review
  of Earth and Planetary Sciences 43 (2015) 509--540.

\bibitem{showman2020atmospheric}
A.~P. Showman, X.~Tan, V.~Parmentier, Atmospheric dynamics of hot giant planets
  and brown dwarfs, Space Science Reviews 216~(8) (2020) 1--83.

\bibitem{perez2013atmospheric}
D.~Perez-Becker, A.~P. Showman, Atmospheric heat redistribution on hot
  jupiters, The Astrophysical Journal 776~(2) (2013) 134.

\bibitem{tan2019atmospheric}
X.~Tan, T.~D. Komacek, The atmospheric circulation of ultra-hot jupiters, The
  Astrophysical Journal 886~(1) (2019) 26.

\bibitem{freedman2008line}
R.~S. Freedman, M.~S. Marley, K.~Lodders, Line and mean opacities for ultracool
  dwarfs and extrasolar planets, The Astrophysical Journal Supplement Series
  174~(2) (2008) 504.

\bibitem{freedman2014gaseous}
R.~S. Freedman, J.~Lustig-Yaeger, J.~J. Fortney, R.~E. Lupu, M.~S. Marley,
  K.~Lodders, Gaseous mean opacities for giant planet and ultracool dwarf
  atmospheres over a range of metallicities and temperatures, The Astrophysical
  Journal Supplement Series 214~(2) (2014) 25.

\bibitem{lupu2014atmospheres}
R.~Lupu, K.~Zahnle, M.~S. Marley, L.~Schaefer, B.~Fegley, C.~Morley, K.~Cahoy,
  R.~Freedman, J.~J. Fortney, The atmospheres of earthlike planets after giant
  impact events, The Astrophysical Journal 784~(1) (2014) 27.

\bibitem{Chandrasekhar}
S.~Chandrasekhar,
  \href{https://books.google.co.in/books?id=CK3HDRwCT5YC}{Radiative Transfer},
  Dover Books on Intermediate and Advanced Mathematics, Dover Publications,
  1960.
\newline\urlprefix\url{https://books.google.co.in/books?id=CK3HDRwCT5YC}

\bibitem{meurer2017sympy}
A.~Meurer, C.~P. Smith, M.~Paprocki, O.~{\v{C}}ert{\'\i}k, S.~B. Kirpichev,
  M.~Rocklin, A.~Kumar, S.~Ivanov, J.~K. Moore, S.~Singh, et~al., Sympy:
  symbolic computing in python, PeerJ Computer Science 3 (2017) e103.

\bibitem{burrows2008theoretical}
A.~Burrows, J.~Budaj, I.~Hubeny, Theoretical spectra and light curves of
  close-in extrasolar giant planets and comparison with data, The Astrophysical
  Journal 678~(2) (2008) 1436.

\bibitem{seager2010exoplanet}
S.~Seager, Exoplanet atmospheres : physical processes, Princeton University
  Press, Princeton, N.J, 2010.

\bibitem{valencia2013bulk}
D.~Valencia, T.~Guillot, V.~Parmentier, R.~S. Freedman, Bulk composition of gj
  1214b and other sub-neptune exoplanets, The Astrophysical Journal 775~(1)
  (2013) 10.

\bibitem{peraiah1973numerical}
A.~Peraiah, I.~Grant, Numerical solution of the radiative transfer equation in
  spherical shells, IMA Journal of Applied Mathematics 12~(1) (1973) 75--90.

\bibitem{peraiah2002introduction}
A.~Peraiah, An Introduction to Radiative Transfer: Methods and applications in
  astrophysics, Cambridge University Press, 2002.

\bibitem{molliere2015model}
P.~Molli{\`e}re, R.~van Boekel, C.~Dullemond, T.~Henning, C.~Mordasini, Model
  atmospheres of irradiated exoplanets: The influence of stellar parameters,
  metallicity, and the c/o ratio, The Astrophysical Journal 813~(1) (2015) 47.

\bibitem{mansfield2018hst}
M.~Mansfield, J.~L. Bean, M.~R. Line, V.~Parmentier, L.~Kreidberg, J.-M.
  D{\'e}sert, J.~J. Fortney, K.~B. Stevenson, J.~Arcangeli, D.~Dragomir, An
  hst/wfc3 thermal emission spectrum of the hot jupiter hat-p-7b, The
  Astronomical Journal 156~(1) (2018) 10.

\bibitem{husser2013new}
T.-O. Husser, S.~Wende-von Berg, S.~Dreizler, D.~Homeier, A.~Reiners,
  T.~Barman, P.~H. Hauschildt, A new extensive library of phoenix stellar
  atmospheres and synthetic spectra, Astronomy \& Astrophysics 553 (2013) A6.

\bibitem{hartman2011hat}
J.~D. Hartman, G.~Bakos, G.~Torres, D.~W. Latham, G.~Kov{\'a}cs, B.~B{\'e}ky,
  S.~Quinn, T.~Mazeh, A.~Shporer, G.~Marcy, et~al., Hat-p-32b and hat-p-33b:
  Two highly inflated hot jupiters transiting high-jitter stars, The
  Astrophysical Journal 742~(1) (2011) 59.

\bibitem{stassun2017accurate}
K.~G. Stassun, K.~A. Collins, B.~S. Gaudi, Accurate empirical radii and masses
  of planets and their host stars with gaia parallaxes, The Astronomical
  Journal 153~(3) (2017) 136.

\bibitem{wong20163}
I.~Wong, H.~A. Knutson, T.~Kataria, N.~K. Lewis, A.~Burrows, J.~J. Fortney,
  J.~Schwartz, A.~Shporer, E.~Agol, N.~B. Cowan, et~al., 3.6 and 4.5 $\mu$m
  spitzer phase curves of the highly irradiated hot jupiters wasp-19b and
  hat-p-7b, The Astrophysical Journal 823~(2) (2016) 122.

\bibitem{machalek2008thermal}
P.~Machalek, P.~R. McCullough, C.~J. Burke, J.~A. Valenti, A.~Burrows, J.~L.
  Hora, Thermal emission of exoplanet xo-1b, The Astrophysical Journal 684~(2)
  (2008) 1427.

\bibitem{goldstein1960infrared}
J.~Goldstein, The infrared reflectivity of a planetary atmosphere., The
  Astrophysical Journal 132 (1960) 473.

\bibitem{bellman1967chandrasekhar}
R.~Bellman, H.~Kagiwada, R.~Kalaba, S.~Ueno, Chandrasekhar's planetary problem
  with internal sources, Icarus 7~(1-3) (1967) 365--371.

\end{thebibliography}

\end{document}